\documentclass[journal]{IEEEtran}
\ifCLASSINFOpdf
\else
\fi
\usepackage{amsmath}
\usepackage[nospace,noadjust]{cite}
\usepackage{amsfonts}
\usepackage{caption}
\usepackage[usenames,dvipsnames,table]{xcolor}
\usepackage{multirow}
\usepackage{booktabs}
\usepackage{pgfplots}
\usepackage{smartdiagram}
\usepackage{metalogo}
\usepackage{balance}
\usepackage[caption=false]{subfig}
\usepgfplotslibrary{units}
\usepgfplotslibrary{groupplots}
\usetikzlibrary{patterns}
\usepackage{tikz}
\usepackage{pgfplots}
\usepackage{algorithm}
\usepackage[noend]{algpseudocode}
\usetikzlibrary{arrows,shapes,backgrounds,fit}
\usepgfplotslibrary{colorbrewer}
\usetikzlibrary{calc,positioning,shapes,decorations.pathreplacing}
\usetikzlibrary{spy}
\usepackage{graphicx}
\DeclareCaptionLabelSeparator{periodspace}{.\quad}
\usepackage[nolist]{acronym}

\usepackage{hyperref}
\pgfplotsset{compat=1.14}

\begin{document}
%
\title{
Crowdsourced wireless spectrum anomaly detection
}
%
%
%
\author{\IEEEauthorblockN{Sreeraj Rajendran\IEEEauthorrefmark{1}, Vincent Lenders\IEEEauthorrefmark{2}, Wannes Meert\IEEEauthorrefmark{3},  and Sofie Pollin\IEEEauthorrefmark{1}}
\\
\IEEEauthorblockA{
Email: \{sreeraj.rajendran, sofie.pollin\}@esat.kuleuven.be\\
wannes.meert@cs.kuleuven.be, vincent.lenders@armasuisse.ch.\\
\IEEEauthorrefmark{1}Department ESAT, KU Leuven, Belgium,
\IEEEauthorrefmark{3}Department of Computer Science, KU Leuven, Belgium\\
\IEEEauthorrefmark{2}armasuisse, Thun, Switzerland}}

\maketitle

\begin{abstract}
Automated wireless spectrum monitoring across frequency, time and space will be essential for many future applications. Manual and fine-grained spectrum analysis is becoming impossible because of the large number of measurement locations and complexity of the spectrum use landscape. Detecting unexpected behaviors in the wireless spectrum from the collected data is a crucial part of this automated monitoring, and the control of detected anomalies is a key functionality to enable interaction between the automated system and the end user. In this paper we look into the wireless spectrum anomaly detection problem for crowdsourced sensors. We first analyze in detail the nature of these anomalies and design effective algorithms to bring the higher dimensional input data to a common feature space across sensors. Anomalies can then be detected as outliers in this feature space. In addition, we investigate the importance of user feedback in the anomaly detection process to improve the performance of unsupervised anomaly detection. Furthermore, schemes for generalizing user feedback across sensors are also developed to close the anomaly detection loop.
\end{abstract}


%
\IEEEpeerreviewmaketitle

\section{Introduction}
%
%
%
%

 

\begin{acronym}[HBCI]
%
%
%
%
%

\acro{3gpp}[3GPP]{3\textsuperscript{rd} Generation Partnership Program}
\acro{cnn}[CNN]{Convolutional Neural Network}
\acro{fbmc}[FBMC]{Filter Bank Multicarrier}
\acro{phy}[PHY]{physical layer}
\acro{pu}[PU]{Primary User}
\acro{rat}[RAT]{Radio Access Technology}
\acro{rfnoc}[RFNoC]{RF Network on Chip}
\acro{sdr}[SDR]{Software Defined Radio}
\acro{su}[SU]{Secondary User}
\acro{toa}[TOA]{Time of Arrival}
\acro{tdoa}[TDOA]{Time Difference of Arrival}
\acro{usrp}[USRP]{Universal Software Radio Peripheral}
\acro{amc}[AMC]{Automatic Modulation Classification}
\acro{lstm}[LSTM]{Long Short Term Memory}
\acro{soa}[SoA]{state-of-the-art}
\acro{fft}[FFT]{Fast Fourier Transform}
\acro{wsn}[WSN]{Wireless Sensor Networks}
\acro{iq}[IQ]{In-phase and quadrature phase}
\acro{snr}[SNR]{signal-to-noise ratio}
\acro{sps}[sps]{samples/symbol}
\acro{awgn}[AWGN]{Additive White Gaussian Noise}
\acro{ofdm}[OFDM]{Orthogonal Frequency Division Multiplexing}
\acro{los}[LOS]{line of sight}
\acro{psd}[PSD]{Power Spectral Density}
\acro{svm}[SVM]{Support Vector Machines}
\acro{aae}[AAE]{Adversarial Autoencoder}
\acro{vae}[VAE]{Variational Autoencoder}
\acro{saif}[SAIFE]{Spectrum Anomaly Detector with Interpretable FEatures}
\acro{roc}[ROC]{Receiver operating characteristic}
\acro{rnn}[RNN]{Recurrent Neural Network}
\acro{auc}[AUC]{Area Under Curve}
\acro{dsa}[DSA]{Dynamic Spectrum Access}
\acro{hmm}[HMM]{Hidden Markov Models}
\acro{iot}[IoT]{Internet of Things}
\acro{api}[API]{Application Programming Interface}
\acro{gnss}[GNSS]{Global Navigation Satellite System}
\acro{cobras}[COBRAS]{Constraint-based Repeated Aggregation and Splitting}
\acro{ssdo}[SSDO]{Semi-Supervised Detection of Outliers}
\acro{osvm}[OSVM]{One class Support Vector Machine}
\acro{ifo}[IFO]{Isolation Forest}
\acro{lof}[LOF]{Local Outlier Factor}
\acro{rcov}[RCOV]{Robust Covariance}
\acro{loda}[LODA]{Lightweight on-line detector of anomalies}
\acro{ari}[ARI]{Adjusted Rand Index}
\end{acronym}
Radio spectrum is one of our most precious and widely used natural resources. With the advent of new wireless communication technologies, spectrum usage has become very complex resulting in airwave congestion and other interference issues \cite{denisowski2011recognizing, viavi_web}. Diverse spectrum regulations across countries have also contributed to this chaotic spectrum usage when non-standardized wireless devices cross country borders. In addition, easily available illegal wireless jammers or low-cost \ac{sdr} devices which are capable of generating custom wireless signals are making the problem worse. Unintentional and intentional jamming of localization services such as \ac{gnss}, which is used for a wide range of applications such as automated vehicle navigation, airplane landing procedures and maritime vessel tracking, is increasing with the availability of easy jamming devices \cite{thombre2018gnss, strike3_web, airtraffic_sec}. Furthermore, illegal repeaters, which are used to boost mobile coverage, can adversely affect the cell planning of mobile operators resulting in poor coverage and dropouts \cite{illegal_repeaters_web}. On the other hand, densified small cells are becoming fundamental for the new high throughput and low latency networks \cite{matinmikko2017micro}. Automated analysis and detection of anomalous behaviors in the spectrum is becoming crucial to promote the growth of new generation wireless systems. 

The wide range of wireless spectrum anomalies makes it infeasible to collect and label all types of anomalies and formulate anomaly detection as a supervised learning problem. A majority of the algorithms in literature \cite{aldoanomaly,anomalycognitive,o2016recurrent, tandiya2018deep, feng2017anomaly, saife} formulate wireless spectrum anomaly detection as an unsupervised learning problem where the models try to learn the normal spectrum data distributions and detect the uncommon patterns as anomalous. These models make the basic assumption that non-frequent behaviour is anomalous which is not always true. For example, transmissions in ISM bands exist that are very sparse but are not anomalies. Similarly, pirate transmitters or transmission duty cycle limit violations in a unlicensed band are anomalous even when they are very frequent. In this paper we solve wireless spectrum anomaly detection by formulating it as a crowdsourced active learning problem. More specifically we provide solutions to the following questions addressing various challenges posed by them:

\begin{enumerate}
\item How to optimize anomaly detection based on user feedback from different users in a crowdsourced network? 
\item How to improve the anomaly detection of a particular user by using expert feedback from other users which may not have fully overlapping interests?
\item  How to generalize user feedback across different sensors?
\end{enumerate}

\begin{figure*}[t]
 \centering
 \includegraphics[width=\linewidth]{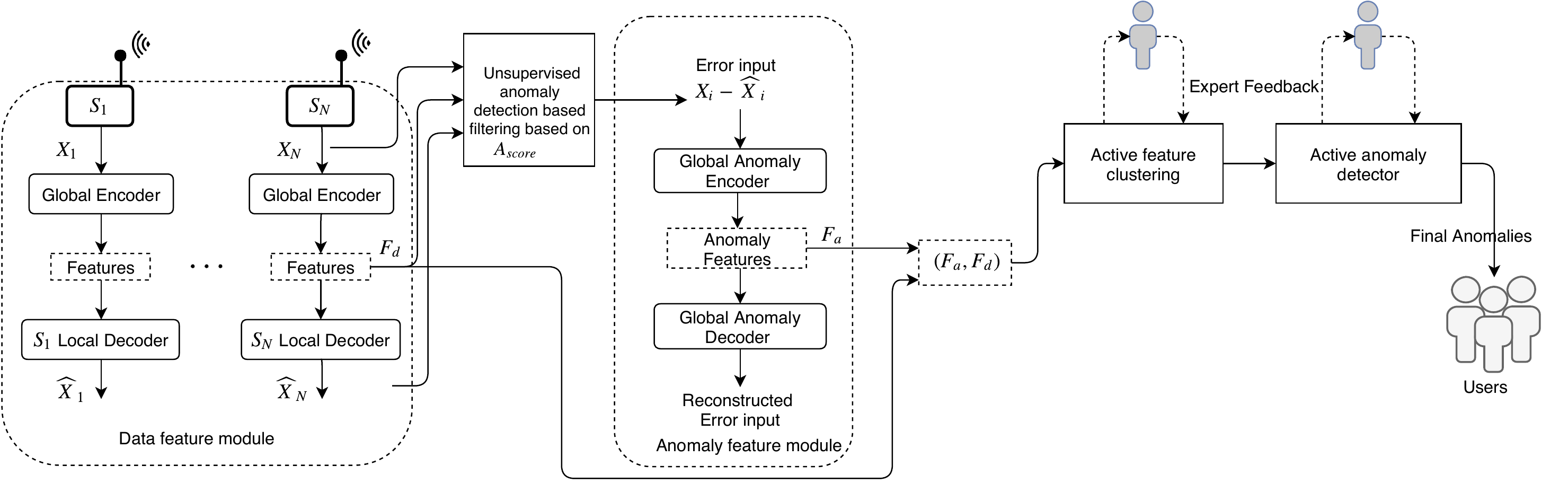}
 \caption{Anomaly detection and clustering framework: The framework consists of an unsupervised data feature module and anomaly feature module for optimized feature extraction from the data collected from multiple sensors $S_1$ to $S_N$. These features are further used for active anomaly clustering and detection with user interaction.}
  \label{fig:anomalyframework} 
\end{figure*}

Some of the major challenges that need to be addressed to answer the aforementioned questions are given below.

\subsection{Challenge 1: Finding an interpretable feature space} 
Contrary to other sensing contexts such as temperature and air pollution monitoring, the amount of data generated by wireless spectrum sensors is quite huge. In addition to the huge amount of data, most of the interesting techniques such as classification, prediction and anomaly detection suffer from the curse of dimensionality when addressed in the higher dimensional input data space. Furthermore, the data costs related to the transfer and storage of the spectrum sensing sensor data is huge. For instance, the crowdsourced spectrum sensing framework, Electrosense \cite{electrosense}, enables three pipelines with very low, medium and high data transfer costs namely: Feature, \ac{psd} and \ac{iq} pipeline respectively. For fulfilling the long term spectrum monitoring dream, it is essential to extract critical features and employ the feature pipeline whenever possible. First, these features should also be compressed and interpretable for further analyzing the cause for the anomaly. Second, the compressed selected feature space should promote outlier detection. Finally, the selected feature space should be the same across sensors for unifying the anomaly detection process for different users. 

\textit{Our contribution:} We introduce novel data driven feature extraction modules for both normal and anomalous wireless spectrum input data. The proposed training process helps to generate a common feature space across sensors without compromising the model's anomaly detection capabilities of individual sensors.

\subsection{Challenge 2: Detecting meaningful anomalies}
Anomalies can be outliers in the normal feature space or process variations. Anomalies can also be some frequent events hidden inside the feature space, such as transmitter duty cycle variations, which prevents us from putting a simple outlier threshold in the feature space. It is also challenging to separate anomalies in an unsupervised feature space across a large number of crowdsourced sensors. The developed framework should also detect a change in the process generating the input spectrum data and retrain the unsupervised model in case of large input distribution variations. 

\textit{Our contribution:} An unsupervised anomaly detection phase followed by a semi-supervised phase is introduced to detect anomalies from frequent as well as less frequent features. Additionally, the proposed framework incorporates provisions for automated model retraining and threshold adjustments of the unsupervised detection phase enabling improved anomaly detection across sensors.

\subsection{Challenge 3: Anomaly knowledge sharing across sensors}
Large variations of sensed wireless spectrum from sensors at different physical locations and regions makes anomaly knowledge sharing across sensors challenging. Knowledge about common anomaly types should be shared across all sensors and frequency bands for fast anomaly detection and generalization of the cause of anomalies in different locations.

\textit{Our contribution:} We introduce an active anomaly clustering process, as shown in Figure~\ref{fig:anomalyframework}, to group similar anomalies across sensors from different frequency bands with varying \ac{snr}s. The specialized active clustering helps to learn relevant local level distances to separate even closely located clusters if they show different behaviour. We show that knowledge acquired through pairwise queries on a single sensor can be effectively used to group similar anomalies from other sensors. 

\subsection{Challenge 4: Incorporating user feedback}
In a crowdsourced network anomalies vary based on user or application perspective. One user's signal might be another's anomaly. So it is important to curate and present anomalies to the users based on their preferences across multiple sensors. In addition, the user anomaly feedback can aid the global anomaly detection process by semi-supervised learning. Thus the main challenge is to device a scheme that can make use of expert labeling from minimal interaction with the users. The labeling process should neither be labour intensive for the users nor prone to much accuracy variations with some level of expected incorrect labeling.

\textit{Our contribution:} An active anomaly detection after interactive clustering is proposed, which makes use of label propagation on the clusters, for anomaly detection. Anomaly detection over clusters helps the framework to learn anomalies from a few labeled instances (as anomalous or not) in each cluster and even easily store anomalies based on user preferences. Furthermore, the proposed anomaly detection model parameters can be adjusted to combat incorrect answering both in the active clustering and anomaly detection phases. 

The rest of the paper is organized as follows. Relevant anomaly detectors proposed in the literature and presented in Section~\ref{related}. Section~\ref{models} explains the anomaly detection framework along with the training and detection algorithms. The dataset used for analyzing the performance of the proposed framework is presented in Section~\ref{dataset}. The performance results and the advantages of the proposed model are discussed in Section~\ref{results}. The paper is concluded in Section~\ref{conclusion}.

\section{Related work}
\label{related}
The anomaly detectors proposed in literature can be mainly brought down to two major classes: unsupervised and semi-supervised anomay detection.

\subsection{Unsupervised anomaly detection}
Various general anomaly detectors are proposed in literature such as \ac{osvm}, \ac{ifo} \cite{liu2012isolation} and \ac{lof} \cite{lof} which makes the basic assumption that outliers represent potential anomalies. These models normally perform poor with increasing input data dimensionality.  An anomaly detector for \ac{dsa} is presented in \cite{aldoanomaly}, where distributed power measurements via cooperative sensing are used for anomaly detection. The proposed detector is limited to authorized user anomaly detection only, for the specific case of \ac{dsa}. Similarly \cite{anomalycognitive} makes use of \ac{hmm} on spectral amplitude probabilities that can detect interference on the channel of interest again in the \ac{dsa} domain. Recently in \cite{o2016recurrent}, the authors presented a \ac{rnn} anomaly detector based on predictive modeling of raw \ac{iq} data. The authors used a \ac{lstm} model for predicting the next 4 \ac{iq} samples from the past 32 samples and an anomaly is detected based on the prediction error. Even though this model works on raw physical layer data which requires no expert feature extraction, it is still not sufficiently automated and generic for practical anomaly detection.  

In \cite{tandiya2018deep}, the authors extend this prediction idea on spectrograms and test the model on some synthetic anomalies. A reconstruction based anomaly detector based on vanilla deep autoencoders is presented in \cite{feng2017anomaly}. In \ac{saif} \cite{saife} we proposed an \ac{aae} \cite{makhzani2015adversarial} based anomaly detector which fills the shortcomings of all the aforementioned methods and can be trained in a semi-supervised fashion for extracting interpretable features, localize anomalies in the spectrum along with data compression and signal classification abilities. An adapted version of \ac{saif} is used for the initial unsupervised anomaly detection part of the proposed framework in this paper.

\subsection{Semi-supervised anomaly detection}
A few semi-supervised anomaly detection algorithms exist in literature such as \ac{loda} \cite{pevny2016loda,fbanomaly}. In \cite{pevny2016loda}, the authors make use of an unsupervised ensemble-based algorithm which is further enhanced with an active learning phase in \cite{fbanomaly}. A semi-supervised accuracy-at-the-top loss function is used to update weights of \ac{loda} based on user feedback. The base anomaly detection algorithm is not limited to \ac{loda} and the authors also show that the active anomaly detection phase can be incorporated with tree based algorithms too \cite{fbtreeanomaly}. On the other hand in \cite{gornitz2013toward}, the authors formulate anomaly detection as an optimization problem and propose a generalization using both labeled and unlabeled instances. Recently, in \cite{vercruyssen2018semi} the authors proposed \ac{ssdo}, a semi-supervised anomaly scoring algorithm which relies on clustering and user labeling.


\section{Anomaly detection framework}
\label{models}
We address all the challenges mentioned in the introduction by developing an unsupervised anomaly detection module followed by  active anomaly clustering and detection modules as shown in the framework Figure~\ref{fig:anomalyframework}. The framework consists of four major components: a data feature module, an anomaly feature module, an active clustering module and an active anomaly detector module which are explained in detail in the following subsections.

\subsection{Module-1: Data feature generator and unsupervised anomaly detection}\label{sec:datafeat}
The data feature module extracts optimized compressed features from the sensed spectrogram data generated by each sensor. We adapt \ac{saif} \cite{saife}, an \ac{aae} based unsupervised anomaly detector, with a different training procedure to address \emph{Challenge 1} mentioned in the introduction. The layers used for the encoder and decoder of both data and anomaly feature modules are summarized in Table~\ref{table:model_layes}. It has been already shown that \ac{saif} achieves very good anomaly detection accuracy when the anomaly is defined as a less frequent event. In addition, \ac{saif} can be trained in a semi-supervised fashion to extract interpretable results such as the signal position and bandwidth. The model can also achieve anomaly localization in the frequency space.

\begin{table}[t]
\begin{center}
\resizebox{\columnwidth}{!}{
\begin{tabular}{|l|l|l|l|l|}
	\hline
    Section  & Layer& Type & Kernel & No. of filters\\
    \hline
    Encoder & 1,2,3 & Conv+maxpool & (3,3) & 32, 32, 32\\
            & 4\_1     & Fully connected (softmax)         & 8 (nlabels) & -\\
            & 4\_2     & Fully connected (linear)         & 50 (nfeatures) & -\\
    \hline
    Decoder & 1,2,3 & Conv+upsample & (3,3) & 32, 64, 64\\
            & 4     & Conv          & (3,3) & 1\\
            & 5     & Fully connected (linear)  & 6*300 & -\\
    \hline
\end{tabular}
}
\end{center}
\caption{Data and anomaly feature module layers.}
\label{table:model_layes}
\end{table}

For this specific crowdsourced problem and further to  help addressing \emph{Challenge 3}, the \ac{saif} encoder and decoder modules are adapted to address the data variability across the sensors. We make use of a global data feature encoder which helps to obtain a \textit{uniform feature space} across sensors for further clustering. In addition, a global encoder also allows to \textit{share knowledge} across sensors and \textit{generalize} the feature space better, as a single sensor might not witness all wireless signals. Along with a common generalized data feature space, the model should also achieve good initial anomaly detection capabilities. To achieve this, we make use of a sensor specific decoder module which prevents the model from averaging out sensor specific anomalies. For example, if we consider FM radio stations at various locations their centre frequencies might be different. If the decoder is trained globally, the model will perceive that an FM transmission is present at all centre frequencies preventing it from detecting an anomalous pirate transmitter at a particular location.

An initial anomaly detection is performed after generating the data features using the standard anomaly detection score \cite{saife} based on the reconstruction loss ($R_l$), discriminator losses ($D_{lcont}$ and $D_{lcat}$) and classification error as summarized below:
\begin{multline}
A_{score} = ( R_l > (\mu_{R_{lt}} + n * \sigma_{R_{lt}}))\\
\vee ( (\mu_{D_{ltcont}} - n * \sigma_{D_{ltcont}}) > D_{lcont} > (\mu_{D_{ltcont}} + n * \sigma_{D_{ltcont}}))\\
\vee ( (\mu_{D_{ltcat}} - n * \sigma_{D_{ltcat}}) > D_{lcat} > (\mu_{D_{ltcat}} + n * \sigma_{D_{ltcat}}))\\
\vee ( Class_{Encoder} != Class_{input})
\end{multline}

A simple n-sigma threshold is employed on the reconstruction and discriminator loss based on the mean ($\mu$) and standard deviation ($\sigma$) values from the training data. An input data frame is classified as anomalous if $A_{score}$ is $True$. The data feature module filters out a large number of normal input spectrum regions enabling the anomaly feature module to learn much more anomaly specific features as shown in Figure~\ref{fig:anomalyframework}.

\subsection{Module-2: Anomaly feature generator}\label{sec:anomalyfeat}
This module captures the anomalous spectrum features that the data feature module is trained not to capture (as the data feature module extracts only frequent data features). The difference between the input and the reconstructed signal ($X-\hat{X}$) of all the anomalies detected by the data feature module is fed to the anomaly feature extractor module. We make use of the difference signal as these would represent only anomalies without normal data. This module is global to all the sensors which creates a common anomaly feature space across sensors for future clustering. An \ac{aae} similar to the one used in the data feature module is also  employed in this module. The whole training process of the data and anomaly feature modules are summarized in Algorithm~\ref{alg:training}. Both the data and anomaly feature use the same encoder and decoder layers as summarized in Table~\ref{table:model_layes}. $Tanh$ activation is used for all convolutional layers and maxpooling and upsampling uses a factor of 2.

\begin{algorithm}[t]
\caption{Data and anomaly features module training }\label{alg:training}
\hspace*{\algorithmicindent} \textbf{Input:} $\mathbf{X}$: a set of spectrograms from different sensors \\
\hspace*{\algorithmicindent} \textbf{Output: } a model that can generate data and anomaly features, $(F_a,F_d)$ 
\begin{algorithmic}[1]
\State Train data feature encoder and decoder to reduce reconstruction loss of $\mathbf{X}$ along with other losses 
\State Global encoder: Freeze encoder weights
\State Sensor $i$ specific decoder $D_i$: Train only the decoder part of the model specific to each sensor $i$ using the data $X_i$ from corresponding sensors.
\State Freeze weights of the data feature module
\State Determine detection threshold $n$, for the $n$-sigma cutoff of \ac{saif}
\State Initial unsupervised anomaly detection with SAIFE
\State Train anomaly feature module on the error signal $X_i-\hat{X_i}$ from all sensors $i$
\State Freeze weights of the anomaly feature module
\end{algorithmic}
\end{algorithm}

\subsection{Module-3: Active clustering module}\label{sec:activeclus}

The extracted anomaly and data features are then fed to an interactive clustering module for further processing. The aim of the clustering module is to group similar anomalies and normal behaviour which are present in different frequency bands across different sensors with minimal user interaction. We make use of the data features along with anomaly features ($F_a,F_d$), to enable users to cluster even anomalies from the normal feature space. \ac{cobras} \cite{van2018cobras}, a query and time efficient interactive clustering algorithm is used for achieving clustering.

\begin{figure}
 \centering
 \includegraphics[width=\linewidth]{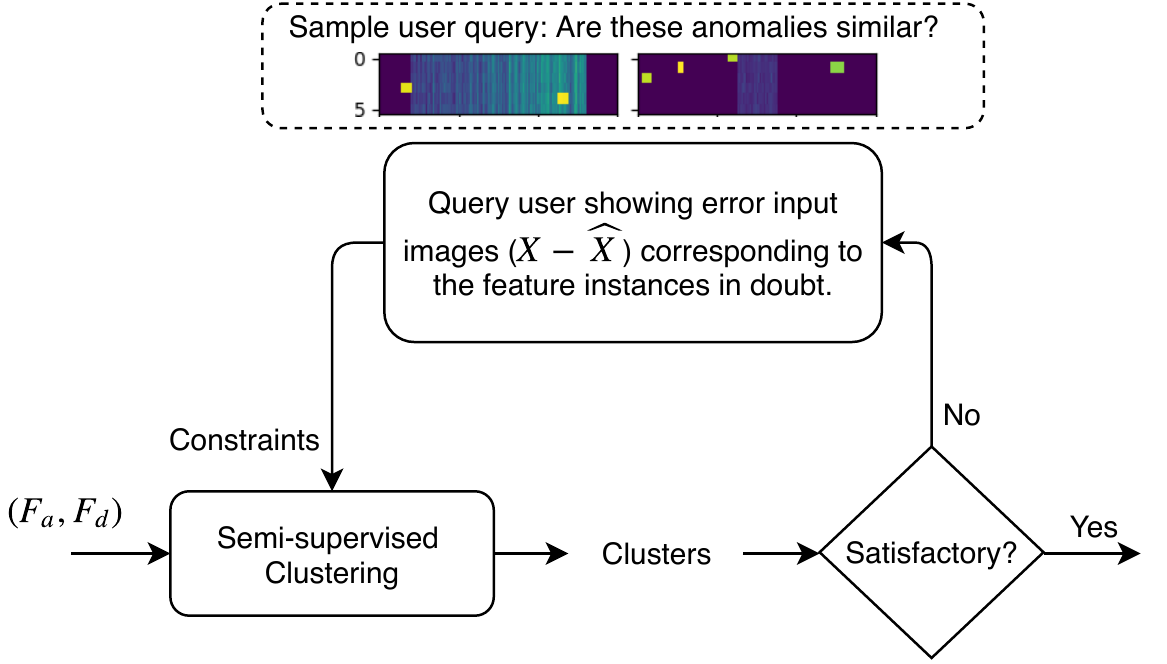}
 \caption[Active clustering]{\ac{cobras} clustering system relying on
interaction in the form of pairwise queries.}
  \label{fig:cobras} 
\end{figure}

\ac{cobras} makes use of the concept of super-instances: sets of instances that are assumed to belong to the same cluster in the unknown target clustering. \ac{cobras} starts with a super-instance that contains all instances and gradually refines the clustering based on pairwise queries asking whether the shown anomalies belong to the same cluster or not. In each iteration \ac{cobras} performs two steps. First, it picks the largest super-instance from the cluster and splits it into new super-instances. A constraint based procedure is used for splitting the largest super-instance. The algorithm splits the largest super-instance into two new temporary super-instances based on 2-means and then queries the user, the relation between their medoids \cite{struyf1997clustering}. Medoids are similar in concept to means, but are restricted to the data points in the dataset or cluster whose average dissimilarity with all the cluster points are minimal. In our case, the error input images ($X_i-\hat{X_i}$) for the corresponding medoid features are shown to the user, asking whether they belong to the same cluster or not, as shown in Figure~\ref{fig:cobras}. In this paper the term \textit{query} when mentioned related to clustering refers to the process of asking the user whether a pair belongs to the same cluster or not. If they can be linked, \ac{cobras} considers the super-instance as pure, otherwise the clustering process is repeated on other super-instances. In the second merging step the relation between newly created clusters are determined and clusters are merged if they belong to the same group based on constraints. If the relation between two clusters is not known, they are queried again for user response. Detailed algorithm steps can be found in the original paper.

\subsection{Module-4: Active anomaly detector}\label{sec:activedetect}

\begin{figure}[t]
\centering
\subfloat[Initial clustering: Intuitive point deviation ($P_d$) and cluster deviation ($C_d$)\label{a}]{
\includegraphics[width=0.6\linewidth]{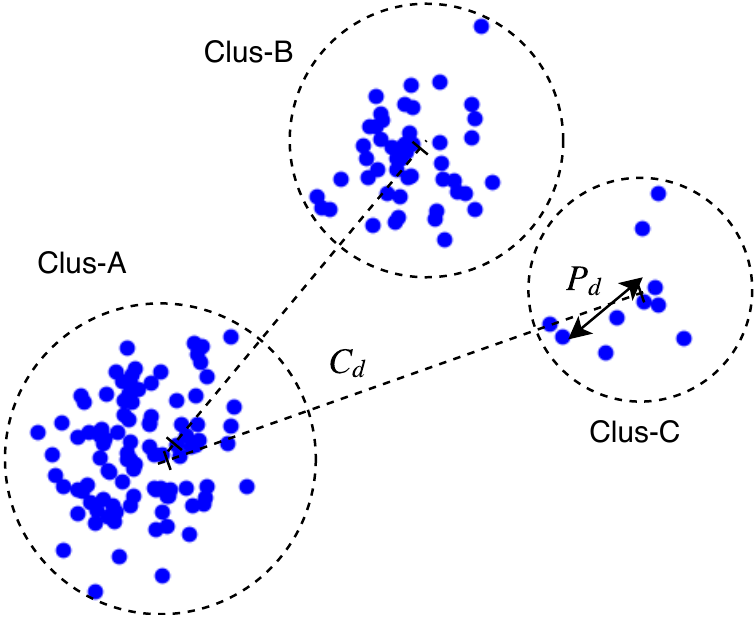}
}\hfill
\subfloat[Unsupervised anomaly detection: detects outliers in clusters.\label{b} ]{
\includegraphics[width=0.47\linewidth]{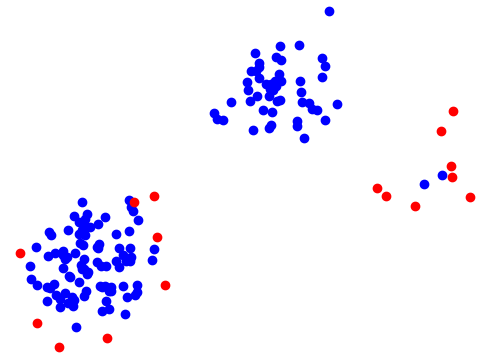}
}\hfill
\subfloat[Anomalies after label propagation: allows to refine anomaly labels for clusters.\label{c}]{
\includegraphics[width=0.46\linewidth]{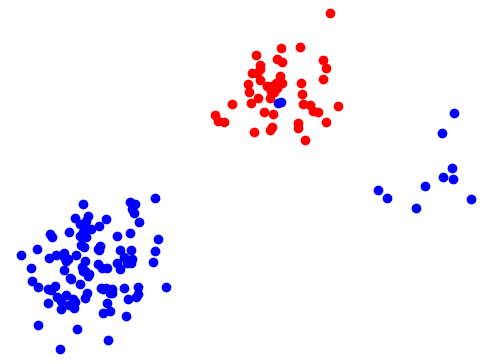}
}\hfill
\caption[SSDO anomaly detection example]{SSDO anomaly detection example: After achieving initial clustering using \ac{cobras}, the algorithm performs unsupervised anomaly scoring and anomaly label propagation. In this example labels are provided saying clusters A and C are normal and Clus-B is anomalous (anomalies in red). }
\label{fig:ssdo}
\end{figure}

The active anomaly detector is an integral module of the entire framework which solves the \emph{Challenge 3} and \emph{Challenge 4} mentioned in the introduction. This module helps to label the clusters generated by \ac{cobras} as anomalous or not. As mentioned in the introduction less frequent events detected by \ac{saif} could be non anomalous. In addition, the active anomaly detection also helps to store user specific anomaly preferences when anomaly labeling is done by a single user, still giving the flexibility to share these anomaly label knowledge across users. For instance, in a crowdsourced network the clusters which are labeled as anomaly by a majority of users can be considered as a global anomaly.

An anomaly score is generated on the clustered instances using \ac{ssdo} \cite{vercruyssen2018semi}, which can work in a semi-supervised setting. \ac{ssdo} performs semi-supervised anomaly detection in two major steps after the clustering step as shown in Figure~\ref{fig:ssdo}. First, it does an unsupervised anomaly scoring of instances based on three measures: (i) the point deviation of an instance from its cluster centre, (ii) the cluster deviation of a selected cluster from other clusters and (iii) the cluster size. In addition, \ac{ssdo} supports score updating based on user anomaly labels with two controllable parameters $\alpha$ and $k$. $\alpha$ controls the impact of the user provided labels versus the labels from the clustering process and $k$ determines the depth of label propagation for a single labeled instance. These parameters are very useful in controlling the speed of the entire process and helps in dealing with noisy labels from users. In this paper the term \textit{query} when mentioned related to anomaly detection refers to the process of asking the user whether an instance is anomalous or not. The whole anomaly detection process is summarized in Algorithm~\ref{alg:anomdetect}. The reader is encouraged to read the original paper \cite{vercruyssen2018semi} for detailed equations of these measures.

\newcommand{\algmargin}{\the\ALG@thistlm}
\makeatother
\newlength{\whilewidth}
\settowidth{\whilewidth}{\algorithmicwhile\ }
\algdef{SE}[parWHILE]{parWhile}{EndparWhile}[1]
  {\parbox[t]{\dimexpr\linewidth-\algmargin}{%
     \hangindent\whilewidth\strut\algorithmicwhile\ #1\ \algorithmicdo\strut}}{\algorithmicend\ \algorithmicwhile}%
\algnewcommand{\parState}[1]{\State%
  \parbox[t]{\dimexpr\linewidth-\algmargin}{\strut #1\strut}}

\begin{algorithm}[t]
\caption{Active Anomaly detection}\label{alg:anomdetect}
\hspace*{\algorithmicindent} \textbf{Input:}  $(F_a,F_d)$ \\
\hspace*{\algorithmicindent} \textbf{Output: } Anomalous spectrograms
\begin{algorithmic}[1]
\State Active clustering on $(F_a,F_d)$ using \ac{cobras}
\State Active anomaly score generation using \ac{ssdo}
\State Present final anomalies to users
\end{algorithmic}
\end{algorithm}

\begin{algorithm}[t]
\caption{Model retraining (Section~\ref{sec:modelretrain})}\label{alg:modelretrain}
\begin{algorithmic}[1]
\State $R_{la}$ = Reconstruction loss of the $X_i-\hat{X_i}$ from the anomaly feature module 
\If{$R_{la}$ $>$ threshold($2\sigma$)} 
\State Add input to the anomaly training set
\EndIf
\While{True}
\State{Retrain anomaly feature module}
\parState{Adjust data feature module detection threshold based on new labeled anomalies not detected by the data feature \ac{saif} and other anomaly clusters}
\parState{Retrain data and anomaly feature module in case of low anomaly scores for majority of clusters}
\State Sleep(threshold\_days)
\EndWhile
\end{algorithmic}
\end{algorithm}

\subsection{Users and expert feedback}
As shown in the framework Figure~\ref{fig:anomalyframework}, users also play an important role in the clustering and anomaly detection process. Electrosense, already leverages crowdsourced sensors for large scale spectrum monitoring. In the newly proposed framework, expert user feedback is utilized both in the clustering and anomaly detection phase. By expert feedback, we refer to feedback from the wireless domain experts: either who develop and train the model or reputed experts among users. As the clustering process helps to share knowledge across sensors, we propose to involve experts in the clustering process.  We also analyze the effect of incorrect labeling or feedback in detail in Section~\ref{results}. The active anomaly detection after clustering could be made open to all users as even with labeling errors we can store them as user specific anomaly preferences and use it globally across sensors based on majority voting. We would like to eventually enhance the spectrum monitoring experience for users with more user engagement in the anomaly detection process, by making this framework a part of Electrosense.

\section{Datasets}
\label{dataset}
Spectrogram data from the \ac{psd} pipeline of four electrosense sensors over a period of 56 days (starting from 8 Nov, 2018)  are collected for the analysis of the proposed framework. The data is collected through the Electrosense API\footnote{\label{noteapi}https://electrosense.org/open-api-spec.html} with a spectral resolution of 10~kHz and time resolution of 60 seconds. The selected sensors are located in Spain, Denmark, Switzerland and Slovenia and data from the same frequency bands are collected from all these sensors as given in Table~\ref{table:dataset_bands}. These sensors are low-cost RTLSDRs configured at a sampling rate of 2.4~MS/s with omni-directional antennas which are deployed indoors. The sensors follow sequential scanning of the spectrum with an \ac{fft} size set to 256 giving a frequency resolution close to 10~kHz. With a \ac{fft} size of 256 and sensor ADC bit-width of 8, we get an effective bitwidth of 12 resulting in a theoretical dynamic range of 74dB. Practical dynamic range depends on the ADC frontend stages and the noise level, which may vary between 60 to 65dB. Five \ac{fft} vectors are averaged for reducing the thermal noise of the receiver.

Three synthetic signals (i) \textit{scont}: single continuous signal with random bandwidth, (ii) \textit{randpulses}: random pulsed transmissions on the given band and (iii) \textit{wpulse}: pulsed wideband signals covering the entire frequency with \ac{snr} from -20 to +20~dB are used as anomalies. These synthetic anomalies are injected to the original collected data which help us to have ground truth clusters (three clusters) and labels of the anomalies for further analysis.

\subsection{Feature extraction model training}
\label{training}
The dataset mentioned in the previous section is split into two subsets, a training and a testing subset, with equal number of vectors. A seed is used to generate random mutually exclusive array indices, which are then used to split the data into two ascertaining the training and testing sets are entirely different. The train and test datasets for performance analysis on the Electrosense dataset are selected from different, non-adjacent, time periods (7 days for training and the other 49 days for testing) to make sure that there are no identical points in the training and test dataset. Spectrogram inputs each with a vector size 6x300 (time-frequency bins) are used as input to the model. The model is trained in an unsupervised fashion to reduce the mean squared error between the input and decoder output and a semi-supervised fashion to learn the the continuous features and class labels. The adversarial networks as well as the autoencoder are trained in three phases: the reconstruction, regularization and semi-supervised phase as mentioned in \cite{makhzani2015adversarial}. The Adam optimizer \cite{adam_optimizer}, a first-order gradient based optimizer, with a learning rate of 0.001 is used for training in all the phases. In the semi-supervised phase the model is trained to learn the class, position and bandwidth of the input signal by training it on 20\% of the labeled samples from the training set. The model is trained for 500 epochs which takes around one hour of training time on a x86 PC with Nvidia GeForce GTX 980 Ti graphics card.

\begin{table}[t]
\begin{center}
\begin{tabular}{|l|l|l|}
	\hline
    Count    & Frequencies (MHz) & Expected technology/modulations\\   						
    \hline
    1 & 86-108 & FM\\
    \hline
    2 & 192-197 & Broadcasting (DAB/Microphones)\\
    \hline
    3 & 198-228 & Broadcasting (DAB/Microphones)\\
    \hline
    4 & 790-801 & Tactical links/TV Broadcasting\\
    \hline
    5 & 801-810 & Tactical links/TV Broadcasting\\
    \hline
    6 & 811-821 & Tactical links/TV Broadcasting\\
    \hline
    7 & 933-935 & Mobile (GSM-900)\\
    \hline
    8 & 955-960 &  Mobile (GSM-900)\\
    \hline
\end{tabular}
\end{center}
\caption{Electrosense dataset frequency bands.}
\label{table:dataset_bands}
\end{table}

\subsection{Model retraining}
\label{sec:modelretrain}
The anomaly and data feature module should be incrementally trained in the following scenarios. An increase in the anomaly detector reconstruction loss ($R_{la}$) indicates that new unseen anomalies are getting fed to the anomaly feature module. This demands retraining of the anomaly feature module. Similarly, a large number of non-anomalous clusters after expert labeling process might either indicate there are large process variations in the sensor output or the selected low frequency events are not anomalous to the users. This demands retraining of the feature extractor modules along with the initial unsupervised threshold update. The framework can also enable users to label portions of the spectrum that are anomalous but not detected by the initial unsupervised detection phase, resulting in a possible threshold reduction for incorporating more normal behaviours to the anomaly space. Summary of the training phase can be found in Algorithms~\ref{alg:training}, \ref{alg:anomdetect} and \ref{alg:modelretrain}.

\section{Results and discussion}
\label{results}
The framework is evaluated based on the unsupervised anomaly detection performance, clustering performance and finally based on the anomaly detection accuracy. We answer the following questions based on the analysis:

\textbf{Q1:} How good is the initial unsupervised anomaly detection performance compared to \ac{soa} models with the new interpretable common feature space and training process (Challenge 1)? 

\textbf{Q2:} Does the common feature space along with active clustering help to share knowledge of anomalies across sensors (Challenge 3)?

\textbf{Q3:} Can the model effectively detect meaningful anomalies even in the feature space with active learning (Challenge 2)? Can the model achieve this with a few labels and work reasonably well even with incorrect feedback (Challenge 4)?

\subsection{\textbf{Q1:} Feature extraction module performance}
\begin{figure}[t]
\centering
\begin{tikzpicture}
\begin{scope}[xshift=0.0cm,yshift=0.0cm]
\begin{axis}[legend style={at={(1.1,1.4)},anchor=north},legend columns=4,width=0.5\columnwidth,grid=both,ylabel= Accuracy(\%),xlabel=SNR,grid style={line width=.1pt, draw=gray!10},major grid style={line width=.2pt,draw=gray!50},xmin=-20,xmax=20,ymax=100, ymin=0,minor tick num=5,legend cell align={left},colormap/hot,,]
\addplot[color= blue,mark size=2pt, mark=square] coordinates 
{
( -20 , 34.04166666666666 )( -10 , 42.66666666666667 )( 0 , 66.16666666666667 )( 10 , 69.00833333333333 )( 20 , 72.775 )
};
\addlegendentry{osvm}
\addplot[color=violet,mark size=3pt, mark=diamond] coordinates {
( -20 , 10.416666666666668 )
( -10 , 16.15833333333333 )
( 0 , 32.391666666666666 )
( 10 , 46.574999999999996 )
( 20 , 54.37500000000001 )
};
\addlegendentry{ifo}

\addplot[color=red, mark size=3pt, mark=triangle] coordinates {
( -20 , 27.325 )
( -10 , 38.74166666666667 )
( 0 , 65.34166666666667 )
( 10 , 81.18333333333332 )
( 20 , 93.14166666666667 )
};\addlegendentry{loda}
\addplot[color=purple,mark size=3pt, mark=oplus] coordinates {
( -20 , 49.566666666666664 )
( -10 , 55.15 )
( 0 , 88.35833333333333 )
( 10 , 93.98333333333334 )
( 20 , 97.95833333333333 )
};
\addlegendentry{saife}
\end{axis}
\end{scope}
\begin{scope}[xshift=3.7cm,yshift=0.0cm]
\begin{axis}[legend style={at={(0.5,1.2)},anchor=north},legend columns=5,width=0.5\columnwidth,grid=both,ylabel=,xlabel=SNR,grid style={line width=.1pt, draw=gray!10},major grid style={line width=.2pt,draw=gray!50},xmin=-20,xmax=20,ymax=15, ymin=0,minor tick num=5,legend cell align={left},colormap/hot,,]
\addplot[color= blue,mark size=2pt, mark=square] coordinates 
{
( -20 , 4.950000000000001 )( -10 , 4.8500000000000005 )( 0 , 4.883333333333333 )( 10 , 4.866666666666667 )( 20 , 4.858333333333333 )
};
\addlegendentry{osvm}
\addplot[color=violet,mark size=3pt, mark=diamond] coordinates {
( -20 , 4.549999999999999 )
( -10 , 4.158333333333333 )
( 0 , 4.608333333333333 )
( 10 , 4.45 )
( 20 , 4.433333333333334 )
};
\addlegendentry{ifo}

\addplot[color=red, mark size=3pt, mark=triangle] coordinates {
( -20 , 2.425 )( -10 , 2.5 )( 0 , 2.3083333333333336 )( 10 , 2.025 )( 20 , 2.5666666666666673 )
};\addlegendentry{loda}
\addplot[color=purple,mark size=3pt, mark=oplus] coordinates {
( -20 , 7.383333333333334 )
( -10 , 7.583333333333334 )
( 0 , 7.7666666666666675 )
( 10 , 7.0249999999999995 )
( 20 , 7.508333333333334 )
};
\addlegendentry{saife}
\legend{};
\end{axis}
\end{scope}
\end{tikzpicture} 
\caption{Averaged true anomaly detection accuracies (left) and false positive rates (right) for different algorithms on 8 frequency bands from four Electrosense sensors. $n$-sigma cutoff $n=4$ for \ac{ifo} and \ac{loda}, $n=6$ for \ac{osvm} and \ac{saif} with $n=2$ for comparison.}
\label{fig:esense_detaccn4}
\end{figure}
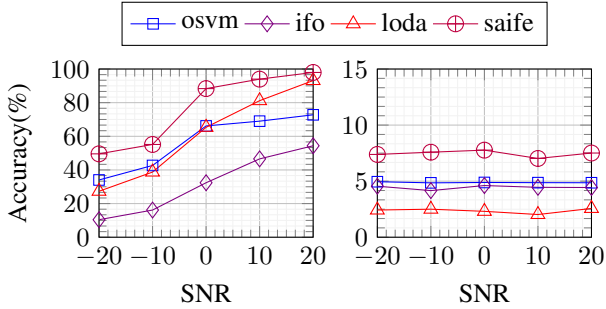

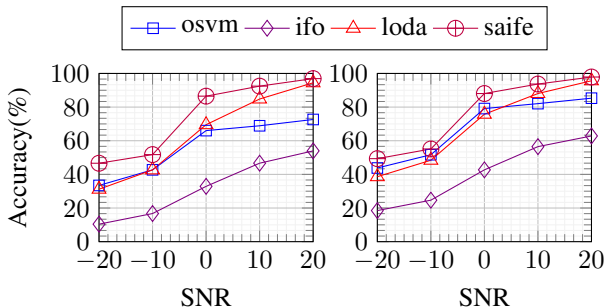
\begin{figure}
\centering
\begin{tikzpicture}
\begin{scope}[xshift=0.0cm,yshift=0.0cm]
\begin{axis}[legend style={at={(1.1,1.4)},anchor=north},legend columns=4,width=0.5\columnwidth,grid=both,ylabel= Accuracy(\%),xlabel=SNR,grid style={line width=.1pt, draw=gray!10},major grid style={line width=.2pt,draw=gray!50},xmin=-20,xmax=20,ymax=100, ymin=0,minor tick num=5,legend cell align={left},colormap/hot,,]
\addplot[color= blue,mark size=2pt, mark=square] coordinates
{
( -20 , 33.425 )
( -10 , 42.69166666666667 )
( 0 , 66.01666666666667 )
( 10 , 68.89166666666667 )
( 20 , 72.63333333333334 )

};
\addlegendentry{osvm}
\addplot[color=violet,mark size=3pt, mark=diamond] coordinates {
( -20 , 10.408333333333335 )
( -10 , 16.71666666666667 )
( 0 , 32.87500000000001 )
( 10 , 46.65 )
( 20 , 53.95833333333334 )
};
\addlegendentry{ifo}

\addplot[color=red, mark size=3pt, mark=triangle] coordinates {
( -20 , 31.183333333333334 )
( -10 , 42.65833333333333 )
( 0 , 69.53333333333335 )
( 10 , 84.8 )
( 20 , 94.45 )

};\addlegendentry{loda}
\addplot[color=purple,mark size=3pt, mark=oplus] coordinates {
( -20 , 46.61666666666666 )
( -10 , 51.75833333333334 )
( 0 , 86.49999999999999 )
( 10 , 92.45833333333331 )
( 20 , 96.93333333333332 )

};
\addlegendentry{saife}
\end{axis}
\end{scope}
\begin{scope}[xshift=3.7cm,yshift=0.0cm]
\begin{axis}[legend style={at={(0.5,1.2)},anchor=north},legend columns=5,width=0.5\columnwidth,grid=both,ylabel=,xlabel=SNR,grid style={line width=.1pt, draw=gray!10},major grid style={line width=.2pt,draw=gray!50},xmin=-20,xmax=20,ymax=100, ymin=0,minor tick num=5,legend cell align={left},colormap/hot,,]
\addplot[color= blue,mark size=2pt, mark=square] coordinates
{
( -20 , 43.78333333333333 )
( -10 , 51.68333333333334 )
( 0 , 79.12500000000001 )
( 10 , 82.05833333333334 )
( 20 , 85.29166666666667 )
};
\addlegendentry{osvm}
\addplot[color=violet,mark size=3pt, mark=diamond] coordinates {
( -20 , 18.474999999999998 )
( -10 , 24.666666666666668 )
( 0 , 42.69166666666667 )
( 10 , 56.49166666666666 )
( 20 , 62.84166666666666 )
};
\addlegendentry{ifo}

\addplot[color=red, mark size=3pt, mark=triangle] coordinates {
( -20 , 38.61666666666666 )
( -10 , 48.4 )
( 0 , 75.75 )
( 10 , 88.05833333333334 )
( 20 , 95.64166666666667 )
};\addlegendentry{loda}
\addplot[color=purple,mark size=3pt, mark=oplus] coordinates {
( -20 , 49.35 )
( -10 , 55.10833333333333 )
( 0 , 88.05833333333334 )
( 10 , 93.69999999999999 )
( 20 , 97.88333333333333 )

};
\addlegendentry{saife}
\legend{};
\end{axis}
\end{scope}
\end{tikzpicture}
\caption{Averaged true anomaly detection accuracies for different algorithms on 8 frequency bands from four Electrosense sensors for a constant false alarm rate of 5\% (left) and 10\%(right).}
\label{fig:esense_detacc_cfar}
\end{figure}

The trained data feature module's performance is compared against various \ac{soa} algorithms such as \ac{osvm}, \ac{ifo} \cite{liu2012isolation} and \ac{loda} \cite{pevny2016loda}. The average anomaly detection accuracy and false positive rates of these algorithms over 8 frequency bands, 3 anomalies and four sensors for a cutoff $n=2$ are plotted in Figure~\ref{fig:esense_detacc}. It can be noticed that \ac{saif} performs better than all algorithms with good detection accuracy and low false positive rates. The false positive rate of other algorithms can be reduced by increasing the $n$ value of the $n$-sigma cutoff threshold as shown in Figure~\ref{fig:esense_detaccn4} at the cost of detection accuracy. These results show that unsupervised threshold selection, even with full anomaly ground truth knowledge, is difficult highlighting the need for a further semi-supervised phase. To systematically compare different algorithms, the detection accuracy for a constant false alarm rate of 5\% and 10\% are plotted in Figure~\ref{fig:esense_detacc_cfar}. It can be noticed that \ac{saif} performs superior to all algorithms. Even with the new training process the performance of the model is consistent with the original model mentioned in \cite{saife} with anomalies like \textit{scont} which simulates new random continuous transmissions in different bands and sensor locations. This shows that the propsoed training process is able to detect new pirate transmitters as discussed in Section~\ref{sec:anomalyfeat}. In addition, \ac{saif} provides an interpretable feature space when compared to other algorithms.

\begin{figure}[t]
\centering
\begin{tikzpicture}
\begin{scope}[xshift=0.0cm,yshift=0.0cm]
\begin{axis}[legend style={at={(1.1,1.4)},anchor=north},legend columns=4,width=0.5\columnwidth,grid=both,ylabel= Accuracy(\%),xlabel=SNR,grid style={line width=.1pt, draw=gray!10},major grid style={line width=.2pt,draw=gray!50},xmin=-20,xmax=20,ymax=100, ymin=0,minor tick num=5,legend cell align={left},colormap/hot,,]
\addplot[color= blue,mark size=2pt, mark=square] coordinates 
{
( -20 , 83.08333333333331 )
( -10 , 87.76666666666667 )
( 0 , 93.35 )
( 10 , 97.625 )
( 20 , 98.19999999999999 )
};
\addlegendentry{osvm}
\addplot[color=violet,mark size=3pt, mark=diamond] coordinates {
( -20 , 54.41666666666666 )
( -10 , 61.30833333333333 )
( 0 , 76.90833333333332 )
( 10 , 83.28333333333333 )
( 20 , 85.74166666666666 )
};
\addlegendentry{ifo}

\addplot[color=red, mark size=3pt, mark=triangle] coordinates {
( -20 , 58.80833333333334 )
( -10 , 65.70833333333333 )
( 0 , 87.35833333333333 )
( 10 , 94.88333333333333 )
( 20 , 98.25 )
};\addlegendentry{loda}
\addplot[color=purple,mark size=3pt, mark=oplus] coordinates {
( -20 , 49.566666666666664 )
( -10 , 55.15 )
( 0 , 88.35833333333333 )
( 10 , 93.98333333333334 )
( 20 , 97.95833333333333 )
};
\addlegendentry{saife}
\end{axis}
\end{scope}
\begin{scope}[xshift=3.7cm,yshift=0.0cm]
\begin{axis}[legend style={at={(0.5,1.2)},anchor=north},legend columns=5,width=0.5\columnwidth,grid=both,ylabel=,xlabel=SNR,grid style={line width=.1pt, draw=gray!10},major grid style={line width=.2pt,draw=gray!50},xmin=-20,xmax=20,ymax=100, ymin=0,minor tick num=5,legend cell align={left},colormap/hot,,]
\addplot[color= blue,mark size=2pt, mark=square] coordinates 
{
( -20 , 74.59166666666667 )
( -10 , 74.825 )
( 0 , 74.72499999999998 )
( 10 , 75.06666666666666 )
( 20 , 75.375 )
};
\addlegendentry{osvm}
\addplot[color=violet,mark size=3pt, mark=diamond] coordinates {
( -20 , 43.63333333333334 )
( -10 , 43.90833333333333 )
( 0 , 43.55833333333334 )
( 10 , 43.53333333333333 )
( 20 , 43.83333333333333 )
};
\addlegendentry{ifo}

\addplot[color=red, mark size=3pt, mark=triangle] coordinates {
( -20 , 34.233333333333334 )
( -10 , 34.741666666666674 )
( 0 , 34.358333333333334 )
( 10 , 34.62499999999999 )
( 20 , 33.78333333333334 )
};\addlegendentry{loda}
\addplot[color=purple,mark size=3pt, mark=oplus] coordinates {
( -20 , 7.383333333333334 )
( -10 , 7.583333333333334 )
( 0 , 7.7666666666666675 )
( 10 , 7.0249999999999995 )
( 20 , 7.508333333333334 )
};
\addlegendentry{saife}
\legend{};
\end{axis}
\end{scope}
\end{tikzpicture} 
\caption{Averaged true anomaly detection accuracies (left) and false positive rates (right) for different algorithms on 8 frequency bands from four Electrosense sensors with $n$-sigma cutoff $n=2$.}
\label{fig:esense_detacc}
\end{figure}
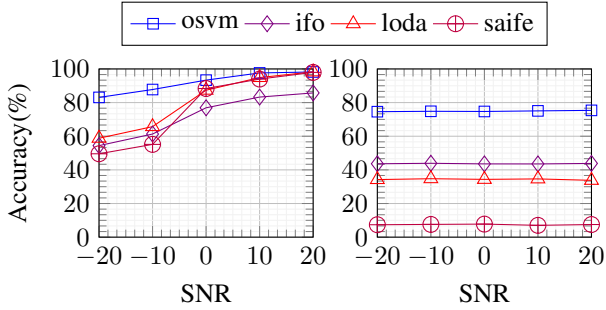

\subsection{\textbf{Q2:} Clustering performance}

\begin{figure}
\centering
\input{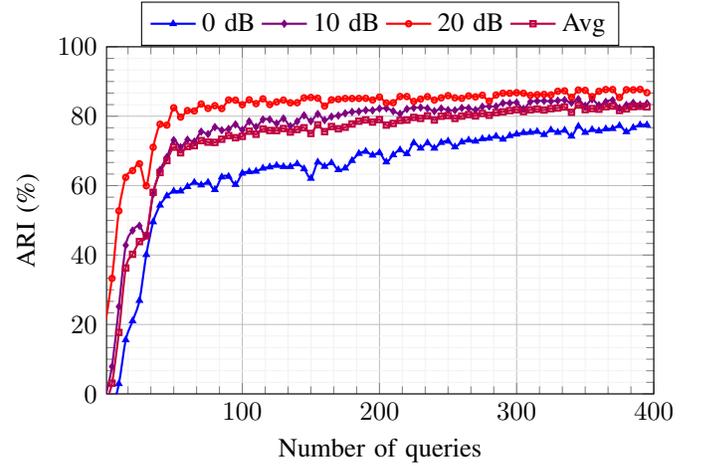} 
\caption{Average clustering ARI with queries answered from all four sensors.}
\label{fig:esense_clusacc}
\end{figure}

\begin{figure}
\centering
\input{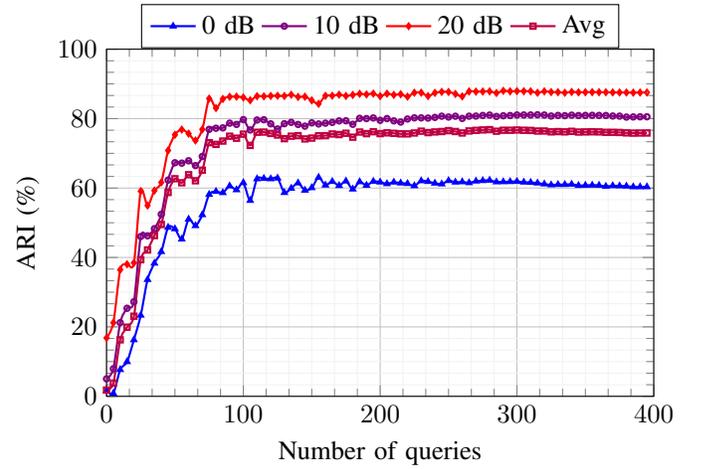} 
\caption{Average clustering ARI with queries answered from only one sensor for different SNRs.}
\label{fig:esense_clusacc_ssens}
\end{figure}

\begin{figure}
\centering
\input{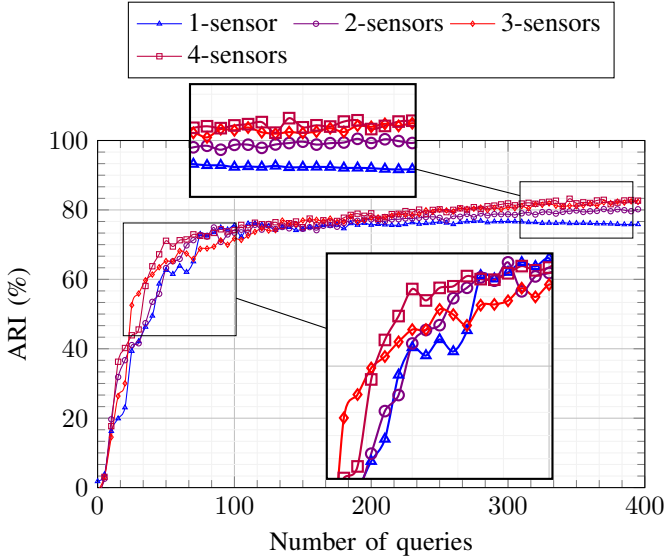} 
\caption{Average clustering ARI with queries answered for different sensor counts.}
\label{fig:esense_clusacc_mulsens}
\end{figure}

\begin{figure}
\centering
\input{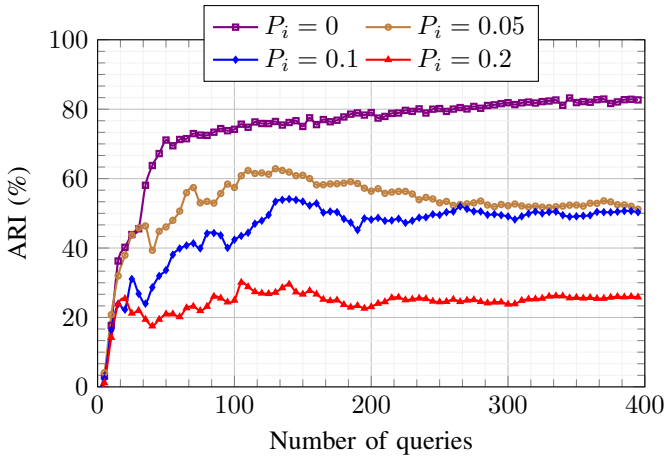} 
\caption{Average clustering ARI with queries answered with different incorrect answering probabilities ($P_i$).}
\label{fig:esense_lprob}
\end{figure}

The average clustering performance of \ac{cobras} for three anomalies (three ground truth clusters) on 8 different frequency bands and four sensors for different \ac{snr}s is plotted in Figure~\ref{fig:esense_clusacc}. The clustering performance measure plotted is \ac{ari}  \cite{hubert1985comparing} which is a symmetric similarity measure between two clusterings, in this case between the produced anomaly clustering and ground truth. As mentioned in Section~\ref{dataset}, we make use of samples with three added synthetic anomalies, providing three clusters as our ground truth. The results shown are averaged scores from 10 runs where \ac{cobras} is allowed to query from data from all sensors. The clustering accuracy of high \ac{snr} anomalies (20 and 10~dB) reach above 80\% with tens of queries while the low \ac{snr} anomalies require queries in hundreds to reach the same. 

In order to test the generalization capability across different sensors the clustering performance is plotted again with only answering queries of anomalies from a single sensor. The results shown in Figure~\ref{fig:esense_clusacc_ssens} confirms that a few more queries are required even at high \ac{snr} anomaly conditions to reach the 80\% mark when compared to answering queries from all sensors as expected.  This validates that the global anomaly feature space across sensors helps to generalize similar anomalies across sensors. After a few queries, the clustering performance saturates and it doesn't keep on improving as in the previous multi-sensor querying scenario. This indeed indicates the need to enable interactive querying from a few different sensors to improve the clustering performance. To further validate this, we include the average \ac{ari} across all SNRs by enabling querying from multiple sensors. The results plotted in Figure~\ref{fig:esense_clusacc_mulsens} shows that the clustering accuracy gets better with a fewer queries with more sensors and it keeps on increasing at a faster rate when compared to lower sensor counts.

Further, the clustering performance is evaluated for understanding the effect of wrong query answers which might happen especially in a crowdsourced expert-feedback paradigm. The average ARI for various probabilities of incorrect answering  ($P_i)$ are shown in Figure~\ref{fig:esense_lprob}. It can be seen that the clustering performance is significantly affected by incorrect answeing. Above $P_i=0.2$ the clustering performance is largely affected. At very high incorrect answering rates it is better to switch to an unsupervised clustering algorithm or adjust the $\alpha$ and $k$ parameters of the anomaly detection algorithm (\ac{ssdo}) to give low priority to the clustering process. 

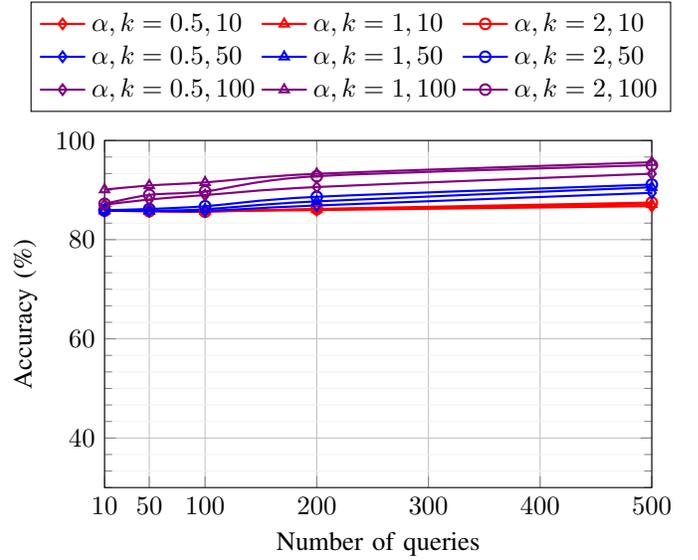
\begin{figure}[t]
\centering
\begin{tikzpicture}
\begin{axis}[legend style={at={(0.45,1.4)},anchor=north},legend columns=3,width=\columnwidth,height=0.7\columnwidth,grid=both,ylabel=Accuracy (\%),xlabel=Number of queries,grid style={line width=.1pt, draw=gray!10},major grid style={line width=.2pt,draw=gray!50},xmin=10,xmax=500,ymax=100, ymin=30,minor tick num=5,legend cell align={left},colormap/hot,xtick={10,50,100,200,300,400,500},]

\addplot[color=red, mark=diamond, mark size=2pt, thick, smooth] coordinates {
( 10 , 85.83 )( 50 , 85.79 )( 100 , 85.81 )( 200 , 85.96 )( 500 , 86.76 )
};
\addlegendentry{$\alpha, k=0.5, 10$}

\addplot[color=red, mark=triangle, mark size=2pt, thick, smooth] coordinates {
( 10 , 85.83 )( 50 , 85.75 )( 100 , 85.73 )( 200 , 86.11 )( 500 , 87.12 )
};
\addlegendentry{$\alpha, k=1, 10$}

\addplot[color=red, mark=o, mark size=2pt, thick, smooth] coordinates {
( 10 , 85.83 )( 50 , 85.74 )( 100 , 85.63 )( 200 , 86.19 )( 500 , 87.48 )
};
\addlegendentry{$\alpha, k=2, 10$}

\addplot[color= blue,mark=diamond, mark size=2pt, thick, smooth] coordinates 
{
( 10 , 85.87 )( 50 , 85.79 )( 100 , 85.79 )( 200 , 86.89 )( 500 , 89.44 )
};
\addlegendentry{$\alpha, k=0.5, 50$}

\addplot[color= blue,mark=triangle, mark size=2pt, thick, smooth] coordinates 
{
( 10 , 85.91 )( 50 , 85.8 )( 100 , 86.13 )( 200 , 87.72 )( 500 , 90.57 )
};
\addlegendentry{$\alpha, k=1, 50$}

\addplot[color= blue,mark=o, mark size=2pt, thick, smooth] coordinates 
{
( 10 , 85.95 )( 50 , 86.18 )( 100 , 86.74 )( 200 , 88.64 )( 500 , 91.11 )
};
\addlegendentry{$\alpha, k=2, 50$}

\addplot[color= violet,mark=diamond, mark size=2pt, thick, smooth] coordinates 
{
( 10 , 87.04 )( 50 , 88.1 )( 100 , 88.99 )( 200 , 90.61 )( 500 , 93.3 )
};
\addlegendentry{$\alpha, k=0.5, 100$}

\addplot[color= violet,mark=triangle, mark size=2pt, thick, smooth] coordinates 
{
( 10 , 90.09 )( 50 , 90.93 )( 100 , 91.55 )( 200 , 93.28 )( 500 , 95.61 )
};
\addlegendentry{$\alpha, k=1, 100$}

\addplot[color= violet,mark=o, mark size=2pt, thick, smooth] coordinates 
{
( 10 , 87.23 )( 50 , 89.03 )( 100 , 89.73 )( 200 , 92.8 )( 500 , 95.02 )
};
\addlegendentry{$\alpha, k=2, 100$}


\end{axis}
\end{tikzpicture} 
\caption{Average anomaly detection accuracy with correct labeling. $\alpha$ controls the impact of the user provided labels versus the labels from the clustering process and $k$ determines the depth of label propagation for a single labeled instance.}
\label{fig:esense_aadacc}
\end{figure}

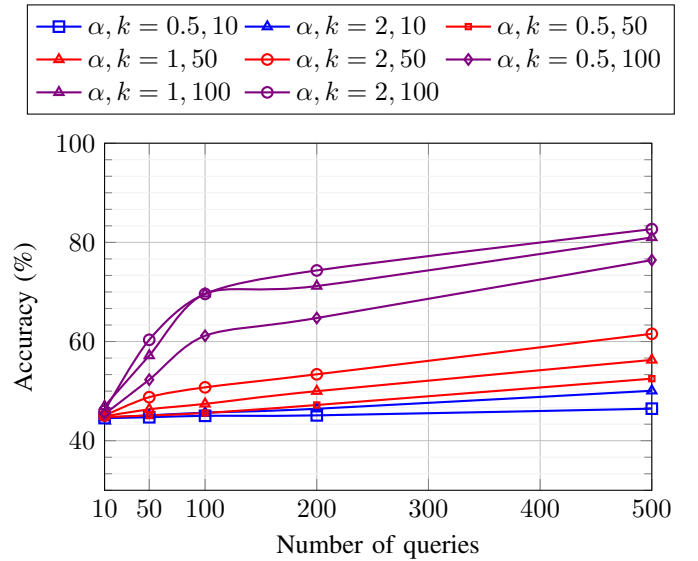
\begin{figure}[t]
\centering
\begin{tikzpicture}
\begin{axis}[legend style={at={(0.45,1.4)},anchor=north},legend columns=3,width=\columnwidth,height=0.7\columnwidth,grid=both,ylabel=Accuracy (\%),xlabel=Number of queries,grid style={line width=.1pt, draw=gray!10},major grid style={line width=.2pt,draw=gray!50},xmin=10,xmax=500,ymax=100, ymin=30,minor tick num=5,legend cell align={left},colormap/hot,xtick={10,50,100,200,300,400,500},]

\addplot[color=blue, mark=square, mark size=2pt, thick, smooth] coordinates {
( 10 , 44.56 )( 50 , 44.73 )( 100 , 45.02 )( 200 , 45.11 )( 500 , 46.46 )
};
\addlegendentry{$\alpha, k=0.5, 10$}

\addplot[color=blue, mark=triangle, mark size=2pt, thick, smooth] coordinates {
( 10 , 44.65 )( 50 , 45.1 )( 100 , 45.66 )( 200 , 46.44 )( 500 , 50.07 )
};
\addlegendentry{$\alpha, k=2, 10$}

\addplot[color=red, mark=square, mark size=1pt, thick, smooth] coordinates {
( 10 , 44.76 )( 50 , 45.09 )( 100 , 45.59 )( 200 , 47.2 )( 500 , 52.51 )
};\addlegendentry{$\alpha, k=0.5, 50$}

\addplot[color=red, mark=triangle, mark size=2pt, thick, smooth] coordinates {
( 10 , 44.9 )( 50 , 46.34 )( 100 , 47.43 )( 200 , 49.99 )( 500 , 56.29 )
};
\addlegendentry{$\alpha, k=1, 50$}

\addplot[color=red, mark=o, mark size=2pt, thick, smooth] coordinates {
( 10 , 45.06 )( 50 , 48.78 )( 100 , 50.76 )( 200 , 53.39 )( 500 , 61.56 )
};
\addlegendentry{$\alpha, k=2, 50$}

\addplot[color= violet,mark=diamond, mark size=2pt, thick, smooth] coordinates 
{
( 10 , 45.37 )( 50 , 52.3 )( 100 , 61.16 )( 200 , 64.74 )( 500 , 76.42 )
};
\addlegendentry{$\alpha, k=0.5, 100$}

\addplot[color=violet, mark=triangle, mark size=2pt, thick, smooth] coordinates {
( 10 , 46.94 )( 50 , 57.17 )( 100 , 69.63 )( 200 , 71.19 )( 500 , 80.99 )
};
\addlegendentry{$\alpha, k=1, 100$}

\addplot[color=violet, mark=o, mark size=2pt, thick, smooth] coordinates {
( 10 , 46.0 )( 50 , 60.35 )( 100 , 69.6 )( 200 , 74.34 )( 500 , 82.67 )
};
\addlegendentry{$\alpha, k=2, 100$}


\end{axis}
\end{tikzpicture} 
\caption{Average anomaly detection accuracy with incorrect cluster and anomaly labeling probability $P_i=0.05$.}
\label{fig:esense_aadacc_fl}
\end{figure}

\subsection{\textbf{Q3:} Anomaly detection performance}
The clustered features are fed to the \ac{ssdo} algorithm for further anomaly labeling as explained in Section~\ref{sec:activedetect}. As the signals fed to the detection phase are initially detected anomalies from the data feature module, we randomly select one of the three synthetic anomalies as non-anomalous and then continue the detection procedure to investigate if we can learn that it is not an anomaly. A random instance is chosen from each cluster and shown to the user for confirming as anomalous or not and the process is continued in a round \textit{robin-fashion} for all clusters. From Figure~\ref{fig:esense_aadacc}, showing the averaged anomaly detection accuracy across all sensors and \ac{snr}s, it is clear that with a few labels or queries itself (each query provides single anomaly label), the model can achieve a good detection accuracy above 90\%. The speed of convergence is also affected by the values of $\alpha$ and $k$, mentioned in the Section~\ref{sec:activedetect}, as expected. The anomaly detection performance is also analyzed for various parameter values with a incorrect labeling probability of $P_i=0.05$ both in the clustering and anomaly detection phase. It can be seen that even with a low starting clustering accuracy of 40\% at $P_i=0.05$ the anomaly detection process is able to reach close to 80\% accuracy with 500 queries. With wrong labeling the convergence is slower which is also an expected behaviour. More advanced strategies such as \textit{selective repeated-labeling} \cite{sheng2008get} can be deployed for improving the detection accuracy under noisy conditions.

\section{Conclusion}
\label{conclusion}
Detecting wireless spectrum anomalies is a challenging problem due to the complexity of wireless spectrum usage and the electromagnetic environment itself. In this paper we have proposed an active framework for wireless spectrum anomaly detection with distributed crowdsourced sensors. The model performs an unsupervised initial anomaly detection and outperforms \ac{soa} algorithms. This unsupervised anomaly detection is followed by a semi-supervised active anomaly detection phase where it provides above 95\% detection accuracy on the selected anomalies. The active anomaly detection phase is also resilient to noisy labeling and it is shown that the model achieves good accuracy close to 80\% with only hundreds of queries. The model also provides good anomaly clustering performance which will be used in future for informing users about relevant anomalies based on user preferences.

\section*{Acknowledgements}
This research was sponsored in part by the NATO Science for Peace and Security Programme under grant G5461.

\appendices




\ifCLASSOPTIONcaptionsoff
  \newpage
\fi



%
\bibliographystyle{IEEEtran}
\bibliography{sections/bibliography}

\begin{IEEEbiography}[{\includegraphics[width=1in,height=1.25in,clip,keepaspectratio]{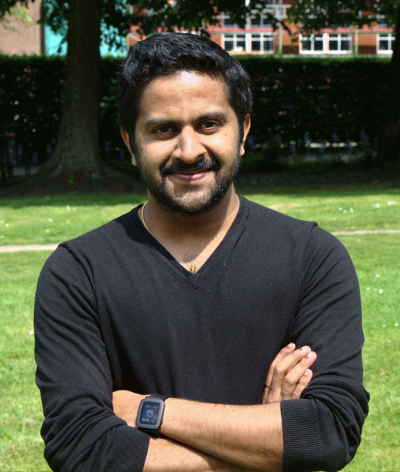}}]
    {Sreeraj Rajendran}
received his Masters degree in communication and signal processing from the Indian Institute of Technology, Bombay, in 2013. He is currently pursuing the PhD degree in the Department of Electrical Engineering, KU Leuven, Belgium. Before joining KU Leuven, he worked as a senior design engineer in the baseband team of Cadence and as an ASIC verification engineer in Wipro Technologies. His main research interests include machine learning algorithms for wireless and low power wireless sensor networks.
\end{IEEEbiography}

\begin{IEEEbiography}[{\includegraphics[width=1in,height=1.25in,clip,keepaspectratio]{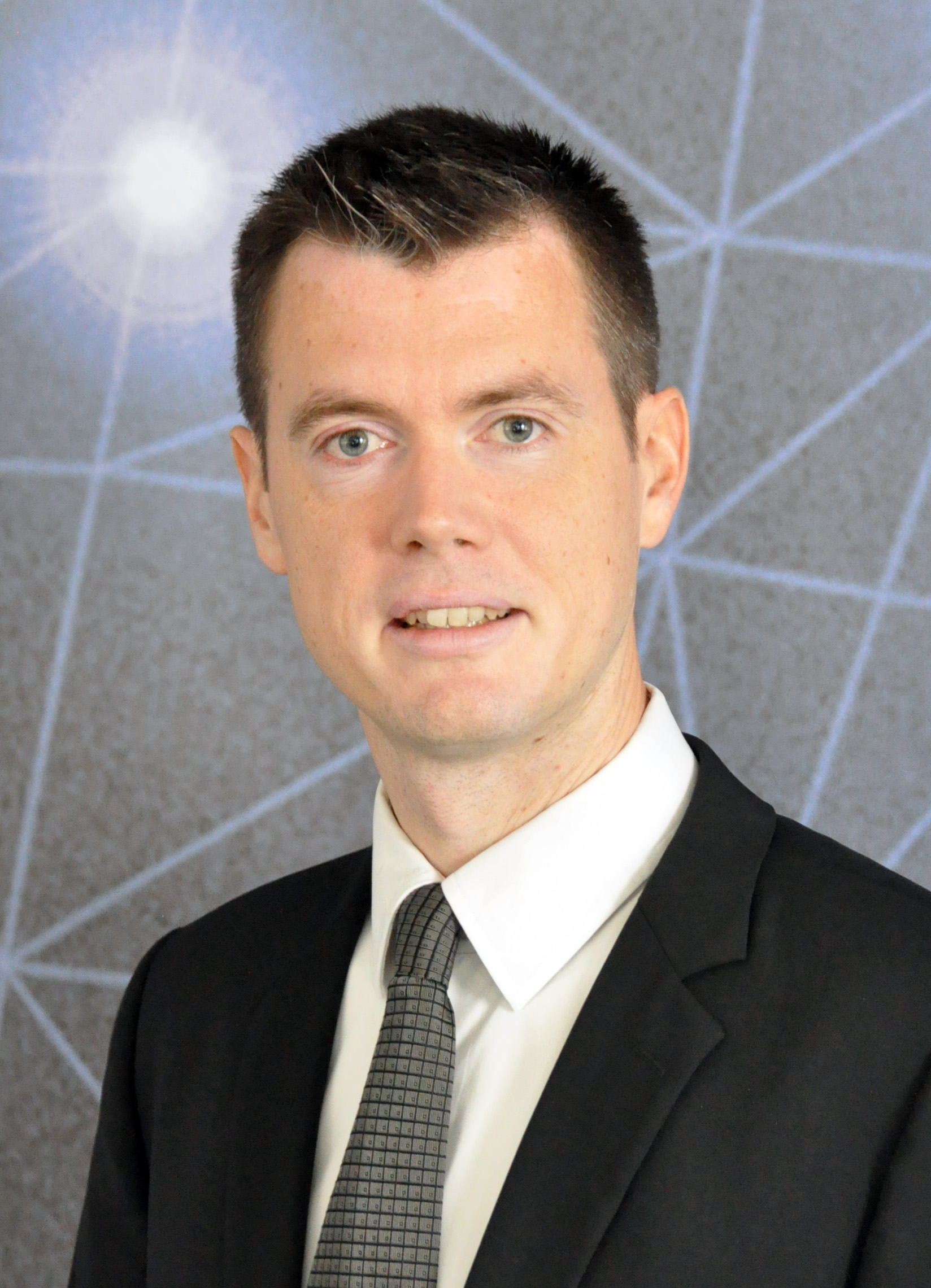}}]
    {Vincent Lenders}
is the head of C4I Networks group and Cyber-Defence Campus at armasuisse. He is also the co-founder and chairman of the executive boards of the OpenSky Network and Electrosense associations. He earned his Ph.D degree (2006) and M.sc (2001) in electrical engineering and information technology both at ETH Zürich, Switzerland. He was also postdoctoral research faculty at Princeton University in the USA.  Vincent Lenders is the author of more than 100 publications that appeared in peer-reviewed international conferences and journals and is the inventor of two patents. He has received best paper awards at IEEE WONS 2012, DFRWS EU 2015, ACM CPSS 2015, and DASC 2015, and the Security Award in 2011 from the Swiss Federal Department of Defense. He is a member of IEEE, ACM, and the expert Jury of the Swiss Economic Forum. He holds various security professional and auditor certifications including CISA, CISM, CRISC (ISACA) and CISSP (ISC2).
\end{IEEEbiography}

\begin{IEEEbiography}[{\includegraphics[width=1in,height=1.25in,clip,keepaspectratio]{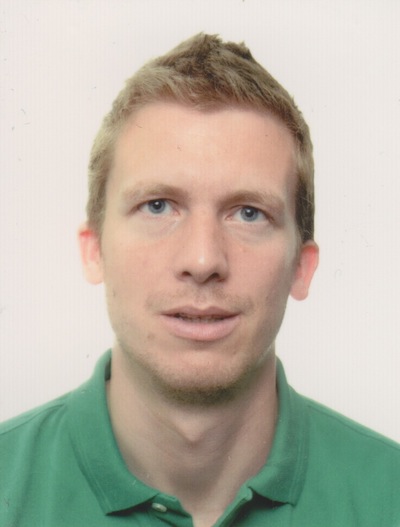}}]
    {Wannes Meert}
received his degrees of Master of Electrotechnical Engineering, Micro-electronics (2005), Master of Artificial Intelligence (2006) and Ph.D. in Computer Science (2011) from KU Leuven. He is currently research manager in the DTAI research group at KU Leuven. His work is focused on applying machine learning, artificial intelligence and anomaly detection technology to industrial application domains.
\end{IEEEbiography}

\begin{IEEEbiography}[{\includegraphics[width=1in,height=1.25in,clip,keepaspectratio]{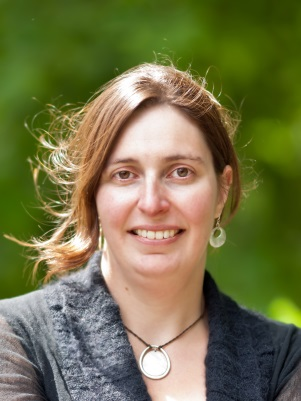}}]
    {Sofie Pollin}
obtained her PhD degree at KU Leuven with honors in 2006. From 2006-2008 she continued her research on wireless communication, energy-efficient networks, cross-layer design, coexistence and cognitive radio at UC Berkeley.  In November 2008 she returned to imec to become a principal scientist in the green radio team. Since 2012, she is tenure track assistant professor at the electrical engineering department at KU Leuven. Her research centers around Networked Systems that require networks that are ever more dense, heterogeneous, battery powered and spectrum constrained. Prof. Pollin is BAEF and Marie Curie fellow, and IEEE senior member. 
\end{IEEEbiography}
%








\end{document}